# A COMPARATIVE STUDY OF SEVERAL PARAMETERIZATIONS FOR SPEAKER RECOGNITION[1]

*Marcos Faúndez-Zanuy*
Escola Universitària Politècnica de Mataró
Universitat Politècnica de Catalunya(UPC)
E08303 MATARO (BARCELONA) SPAIN
Tel: (34) 93 757 44 04 Fax: (34) 93 757 05 24
e-mail: faundez@eupmt.es, http://www.eupmt.es/veu

## ABSTRACT

This paper presents an exhaustive study about the robustness of several parameterizations, in speaker verification and identification tasks. We have studied several mismatch conditions: different recording sessions, microphones, and different languages (it has been obtained from a bilingual set of speakers). This study reveals that the combination of several parameterizations can improve the robustness in all the scenarios for both tasks, identification and verification. In addition, two different methods have been evaluated: vector quantization, and covariance matrices with an arithmetic-harmonic sphericity measure.

## 1. INTRODUCTION

Speech variability is a main degradation factor in speaker recognition tasks. For this reason, it is important to test the speaker recognition algorithms in a wide range of situations, that can be found in a more realistic situation than the laboratory conditions. This paper presents additional results to [1].

### 1.1 Database

The design of the speech corpus and its phonological and syllabic balance follows the parameters proposed in a first database [2]. The main relevant characteristics of our new recordings are the following:

1. 48 bilingual speakers (24 males & 24 females). The speech signal has been acquired at a sampling rate of 16KHz, and all the database is about 20 CD.
2. Four different recording sessions: S1 (first session), S2 (second session, recorded one week later), S3 (recorded another week later) and S4 (1 month after session S3). All the tasks have been sequentially collected in two languages (Catalan and Spanish) uttered from the same speaker This has been done in all the session, so there are S1s, S1c, S2s. S2c, S3s, S3c, S4s, S4c (s=Spanish, c=Catalan).
3. Several tasks have been recorded in each session, including digits, sentences, text, etc. (as described in [2]). These tasks include specific text, different for each speaker, and common text for all the speakers.
4. Each task has been simultaneously recorded with two different microphones: AKG C420, AKG D40S for sessions 1 & 2, AKG C420, SONY ECM-66B for sessions 3 & 4, using one stereo channel for each microphone. In this paper we will use the following notation: M1=AKG C420, M2= AKG D40S, M3=SONY ECM 66B
5. Different recording conditions for all the tasks: One recording session in an anechoic room (S1AR=session one anechoic room), one recording session with a telephone handset plug into a PC connected to an ISDN (S3ISDN Session 3 ISDN recording).

### 1.2 Conditions of the experiments

The speech material has been downsampled to 8KHz. Pre-emphasis of 0.95 was applied. Frames of 240 samples have been chosen (Hamming window), and an overlap between adjacent frames of 2/3. Frames under an energy threshold have been discarded. LPC coefficients have been obtained from each frame, using the Levinson-Durbin recursion. From these coefficients a recursion has been applied in order to obtain the LPCC.

The model of each speaker has been computed using 1 minute of read text (the same text for all the speakers), and the test has been done with 5 sentences for each speaker (each sentence about 2 seconds). Thus, we have made 48x48x5 tests for each parameterization.

Two speaker identification methods have been used in our experiments:
1. Vector quantization [6] (VQ) with a random method for codebook generation (1 codebook for each speaker)
2. Arithmetic-harmonic sphericity measure [7], which implies the computation of a covariance matrix (CM) for each speaker, and the following measure distance:

$$m(C_j C_{test}) = \log\left[tr(C_{test} C_j^{-1}) tr(C_j C_{test}^{-1})\right] - 2\log(P)$$

[1] This work has been supported by the CICYT TIC97-1001-C02-02

For the VQ models, more parameters imply a bigger codebook, while for the CM implies a higher dimensional cepstral vectors.

We have used the following LPCC vectors:

a) $P=16$ for VQ method (LPCC vectors of dimension 16). The codebook size is fixed to $No=6$ bits.

b) $P=20$ for CM (matrices of size 20x20)

In this situation the number of parameters of the VQ method is five times greater than the number of parameters of the CM method.

In both cases $LPCC_{3..P}$ means that the first and second coefficients have been removed.

## 2. STUDIED PARAMETERIZATIONS AND RECOGNITION ALGORITHMS

The studied parameterizations are:

1. Cepstrum (LPCC)
2. Cepstral mean subtraction [3] (CMS)
3. Adaptive component weighted Cepstrum [3] (ACW-LPCC)
4. Cepstral linear weighting [3] (LW-LPCC)
5. Bandpass liftered cepstrum [3] (BPL-LPCC)
6. Cepstral standard deviation weighting ($\sigma$-LPCC) [4]
7. Postfilter Cepstrum [3] (PF-LPCC), with $\alpha=1$, $\beta=0.9$

A combination between several of these parameterizations has also been tested. More information about the way we have used these parameterizations can be obtained in [1]

## 3. RESULTS

We have studied three relevant mismatch training and testing conditions: different microphones, temporal interval between training and recording sessions, and different languages. The best results of each column have been underlined.

### 3.1 Mismatch between training and testing microphones

In this section, the recording session and language are fixed (session 4 Catalan). We evaluate all the possible microphone combinations. For instance, M1M3 means training with microphone M1 and testing with microphone M3. Table 1 summarizes the identification rates using different microphones for S4c (session 4 Catalan), for VQ, as function of the training and testing microphones. Table 2 summarizes the results using the CM in the same conditions as table 1. Table 3 shows the Equal Error Rate (EER) with and without cohorts [7], in a speaker verification task.

| PARAMETERIZ. | M1M1 | M1M3 | M3M3 | M3M1 |
|---|---|---|---|---|
| LPCC | 99,2 | 31,7 | 98,3 | 22,9 |
| $LPCC_{3..P}$ | 95,8 | 39,2 | 97,1 | 24,2 |
| $\sigma$-LPCC | 95,8 | 32,1 | 92,9 | 22,5 |
| ACW | 99,6 | 65,8 | 97,2 | 27,9 |
| CMS | 93,3 | 77,9 | 97,9 | 57,9 |
| CMS+ACW | 95,8 | 87,2 | 97,5 | 76,3 |
| CMS+ACW+$\sigma$-LPCC | 97,5 | 88,8 | 97,9 | 73,3 |
| CMS+$\sigma$-LPCC | 97,5 | 84,6 | 98,3 | 62,6 |
| CMS-LW | 97,1 | 85,5 | 97 | 59,6 |
| ACW+$\sigma$-LPCC | 96,3 | 33,8 | 94,6 | 25 |
| PF | 99,6 | 47,9 | 98,8 | 25,4 |
| CMS+PF | 97,1 | 87,5 | 97,5 | 63,8 |
| CMS+PF+$\sigma$-LPCC | 97,5 | 85 | 98,3 | 63,3 |

*Table 1 Identification rates for session 4 Catalan, VQ*

| PARAMETERIZ. | M1M1 | M1M3 | M3M3 | M3M1 |
|---|---|---|---|---|
| LPCC | 97,1 | 52,5 | 96,3 | 26,3 |
| $LPCC_{3..P}$ | 91,7 | 53,8 | 94,2 | 27,1 |
| $\sigma$-LPCC | 93,8 | 39,6 | 96,7 | 22,9 |
| ACW | 96,7 | 53,8 | 96,3 | 26,7 |
| CMS | 94,2 | 82,1 | 97,2 | 48,3 |
| CMS+ACW | 95,4 | 85,4 | 97,5 | 57,5 |
| CMS+ACW+$\sigma$-LPCC | 95 | 79,6 | 97,9 | 53,8 |
| CMS+$\sigma$-LPCC | 94,2 | 77,1 | 97,1 | 48,8 |
| CMS-LW | 94,2 | 82,1 | 97,9 | 48,3 |
| ACW+$\sigma$-LPCC | 94,6 | 40,4 | 97,5 | 23,3 |
| PF | 97,1 | 52,1 | 96,3 | 26,3 |
| CMS+PF | 94,2 | 82,1 | 97,9 | 48,3 |
| CMS+PF+$\sigma$-LPCC | 94,1 | 77,1 | 97,1 | 48,1 |

*Table 2: Identification rates for session 4 Catalan, CM*

| PARAMETERIZ. | M1M1 | | M1M3 | | M3M3 | | M3M1 | |
|---|---|---|---|---|---|---|---|---|
| LPCC | 0.20 | 1.06 | 10.97 | 10.25 | 0.63 | 0.82 | 22.49 | 18.33 |
| $LPCC_{3..P}$ | 1.50 | 2.32 | 10.77 | 11.57 | 1.02 | 0.38 | 21.34 | 16.71 |
| $\sigma$-LPCC | 1.17 | 1.68 | 15.45 | 17.45 | 0.95 | 0.99 | 27.43 | 24.41 |
| ACW | 0.56 | 1.25 | 9.06 | 8.80 | 0.59 | 0.56 | 20.53 | 15.94 |
| CMS | 1.08 | 1.47 | 4.20 | 3.30 | 0.35 | 0.97 | 9.61 | 6.80 |
| CMS+ACW | 0.85 | 1.40 | 3.28 | 3.10 | 0.27 | 0.90 | 6.68 | 5.11 |
| CMS+ACW+$\sigma$-LPCC | 0.90 | 1.31 | 5.28 | 3.85 | 0.45 | 0.90 | 7.97 | 5.66 |
| CMS+$\sigma$-LPCC | 0.90 | 1.70 | 5.70 | 3.48 | 0.73 | 0.97 | 10.56 | 7.50 |
| CMS-LW | 1.08 | 1.47 | 4.20 | 3.30 | 0.35 | 0.97 | 9.61 | 6.80 |
| ACW+$\sigma$-LPCC | 1.08 | 1.47 | 15.40 | 16.61 | 0.90 | 1.02 | 26.64 | 23.46 |
| PF | 0.20 | 1.06 | 10.57 | 9.66 | 0.67 | 0.84 | 22.51 | 18.12 |
| CMS+PF | 1.08 | 1.47 | 3.76 | 3.32 | 0.56 | 0.97 | 9.61 | 6.83 |
| CMS+PF+$\sigma^2$-LPCC | 0.90 | 1.70 | 5.59 | 3.91 | 0.70 | 0.97 | 10.56 | 7.50 |

*Table 3: EER (%) (with cohorts = 5 / without) for session 4 Catalan.*

### 3.2 Mismatch between training and testing recording sessions.

In this section we have evaluated the relevance of different training and testing recording sessions, and the simultaneous mismatch of microphone and recording session.

Table 4 shows the identification rates in several conditions. For instance, S4cM1S2cM2 means that session 4 Catalan and microphone 1 are used for training, and Session 2 Catalan and microphone 2 for testing. Table 5 summarizes the results using the CM.

| PARAMETERIZ. | S4cM1 S3cM1 | S4cM1 S2cM1 | S4cM1 S2cM2 | S2cM1 S4cM3 |
|---|---|---|---|---|
| LPCC | 85.8 | 75 | 56.7 | 36.3 |
| LPCC$_{3..P}$ | 87.9 | 80 | 73.3 | 39.2 |
| $\sigma$-LPCC | 83.3 | 77,1 | 59.2 | 28.3 |
| ACW | 91.7 | 82,9 | 74.6 | 62.5 |
| CMS | 78.8 | 71,7 | 55.8 | 70.4 |
| CMS+ACW | 90.8 | 81,3 | 77.1 | 67.1 |
| CMS+ACW+$\sigma$-LPCC | 89.2 | 81,7 | <u>77.5</u> | 72.5 |
| CMS+$\sigma$-LPCC | 87.9 | 80,4 | 74.2 | <u>73.3</u> |
| CMS-LW | 87.1 | 79,6 | 75.4 | 68.8 |
| ACW+$\sigma$-LPCC | 83.8 | 79,2 | 62.9 | 29.6 |
| PF | <u>92.5</u> | <u>84,2</u> | 73.4 | 51.7 |
| CMS+PF | 87.1 | 82,5 | 76.3 | 72.9 |
| CMS+PF+$\sigma$-LPCC | 88.3 | 80,8 | 74.2 | 74.2 |

*Table 4 Identification rates for sessions 2,3 & 4 Catalan, VQ (different training and testing sessions).*

| PARAMETERIZ. | S4cM1 S3cM1 | S4cM1 S2cM1 | S4cM1 S2cM2 | S2cM1 S4cM3 |
|---|---|---|---|---|
| LPCC | 84.6 | 73.3 | 62.1 | 48.3 |
| LPCC$_{3..P}$ | 77.5 | 64.2 | 55.8 | 48.8 |
| $\sigma$-LPCC | 77.1 | 70 | 50.4 | 29.6 |
| ACW | <u>85</u> | 74.2 | 63.8 | 48.3 |
| CMS | 821 | 75.8 | 65.8 | 70 |
| CMS+ACW | 84.2 | <u>77.1</u> | <u>67.9</u> | <u>70.4</u> |
| CMS+ACW+$\sigma$-LPCC | 81.7 | 75 | 64.2 | 69.6 |
| CMS+$\sigma$-LPCC | 78.8 | 75.4 | 62.9 | 68.6 |
| CMS-LW | 82.1 | 75.8 | 65.8 | 70 |
| ACW+$\sigma$-LPCC | 79.8 | 71.3 | 49.2 | 33.8 |
| PF | 84.6 | 73.3 | 62.1 | 47.1 |
| CMS+PF | 82.1 | 75.8 | 65.8 | 69.6 |
| CMS+PF+$\sigma$-LPCC | 78.6 | 75.4 | 62.9 | 68.3 |

*Table 5 Identification rates for sessions 2,3 & 4 Catalan, CM (different training and testing sessions).*

| PARAMETERIZ. | S4cM1 S3cM1 | S4cM1 S2cM1 | S4cM1 S2cM2 | S2cM1 S4cM3 |
|---|---|---|---|---|
| LPCC | 3.71 2.69 | 5.30 3.00 | 8.21 8.05 | 12.25 11.02 |
| LPCC$_{3..P}$ | 3.59 3.88 | 7.44 6.06 | 9.98 10.82 | 12.63 11.79 |
| $\sigma$-LPCC | 4.36 <u>2.62</u> | 6.20 3.67 | 9.65 9.13 | 17.75 16.19 |
| ACW | 2.69 2.67 | 5.23 <u>2.77</u> | 7.78 7.15 | 12.16 10.61 |
| CMS | 2.98 3.53 | 5.11 3.13 | 6.97 6.17 | 5.54 5.27 |
| CMS+ACW | <u>2.15</u> 3.46 | 5.00 3.41 | <u>5.70</u> <u>5.40</u> | <u>4.88</u> 5.46 |
| CMS+ACW+$\sigma$-LPCC | 3.40 3.23 | <u>4.77</u> 3.43 | 6.50 6.65 | 6.15 4.85 |
| CMS+$\sigma$-LPCC | 3.38 3.28 | 5.46 2.78 | 6.70 7.58 | 6.83 4.81 |
| CMS-LW | 2.98 3.53 | 5.11 3.13 | 6.97 6.17 | 5.54 5.27 |
| ACW+$\sigma$-LPCC | 3.67 2.85 | 6.88 3.75 | 9.64 8.79 | 17.52 16.27 |
| PF | 3.71 2.69 | 5.30 3.00 | 8.21 8.05 | 11.84 11.59 |
| CMS+PF | 2.98 3.53 | 5.11 3.13 | 6.97 6.17 | 5.35 5.23 |
| CMS+PF+$\sigma$-LPCC | 3.38 3.28 | 5.46 2.78 | 6.70 7.58 | 6.78 <u>4.56</u> |

*Table 6: EER (%) (with cohorts = 5 / without) for sessions 2,3 & 4 Catalan, (different training and testing sessions).*

Obviously in real applications the training and testing sessions are not in the same day, so the identification rates can be degraded. The use of a robust parameterization can improve the results up to 9.2% (same microphone for testing and training), and up to 21% if there is also a change of microphone. Another approach consists on using several recording sessions for training the model of each speaker, instead of using only one recording session, but we have not evaluated this possibility in this study. Table 6 summarizes the verification results for CM method.

### 3.3 Mismatch training and testing recording languages.

This section presents the most relevant results in the following mismatch conditions: a) Training and testing languages, b) training and testing recording languages and microphones, c) Training and testing languages, microphones, and recording sessions.

Table 7 summarizes the identification rates using different languages for training and testing (Catalan=c, Spanish=s), different microphones, and different sessions, using VQ.

Table 8 summarizes the results using the CM. From this table it can be deduced that the change of recording language using the same microphone and recording session has a minor effect over the identification rates. This is because in a speaker identification system the goal is to model the characteristics of the speech production system, not the content of the message. In a similar way, persons can identify familiar voices although they are speaking in an unknown language. On the other hand, one language obtains better results than the other does, but there are small differences. More information about language mismatch can be found in [8].

| PARAMETERIZ. | S4cM1 S4sM1 | S4sM1 S4cM1 | S4cM1 S4sM3 | S2cM1 S4sM3 |
|---|---|---|---|---|
| LPCC | 97.9 | <u>98.8</u> | 31.7 | 29.6 |
| LPCC$_{3..P}$ | 93.8 | 94.6 | 33.3 | 30.8 |
| $\sigma$-LPCC | 90.4 | 97.5 | 32.1 | 19.2 |
| ACW | <u>98.8</u> | <u>98.8</u> | 54.6 | 52.1 |
| CMS | 91.2 | 97.9 | 72.9 | 62.1 |
| CMS+ACW | 96.3 | 95 | <u>88.3</u> | 67.5 |
| CMS+ACW+$\sigma$-LPCC | 96.7 | 97.5 | 85.4 | 68.3 |
| CMS+$\sigma$-LPCC | 94.2 | 96.3 | 81.3 | 69.6 |
| CMS-LW | 96.3 | 97.9 | 81.3 | 63.8 |
| ACW+$\sigma$-LPCC | 90.4 | 95 | 30 | 21.3 |
| PF | <u>98.8</u> | <u>98.8</u> | 41.3 | 39.6 |
| CMS+PF | 95.8 | 97.1 | 86.7 | 69.6 |
| CMS+PF+$\sigma$-LPCC | 90.4 | 97.9 | 81.3 | <u>70</u> |

*Table 7 Identification rates for different training and testing languages, VQ.*

Tables 1 to 9 show that there are small differences in recognition rates between Covariance Matrices and vector quantization using LPCC parameterization, although in the case of mismatch training and testing languages the difference is 4.6% (compare the cells of the first number of tables 7 and 8). VQ outperforms CM for the optimal preprocessing. On the other hand VQ implies more parameters for modeling a given speaker.

| PARAMETERIZ. | S4sM1 S4cM1 | S4cM1 S4sM1 | S2sM1 S4cM3 | S2cM2 S4sM3 |
|---|---|---|---|---|
| LPCC | 93.3 | 97.1 | 45 | 37.5 |
| LPCC$_{3..P}$ | 94.6 | 95 | 52.9 | 40.4 |
| σ-LPCC | 88.8 | 95.4 | 35.8 | 27.1 |
| ACW | 92.1 | 97.5 | 44.6 | 38.3 |
| CMS | 93.3 | 92.5 | 75.4 | 59.2 |
| CMS+ACW | 94.2 | 93.8 | 78.3 | 62.9 |
| CMS+ACW+σ-LPCC | 93.3 | 94.2 | 75 | 58.8 |
| CMS+σ-LPCC | 92.9 | 93.8 | 72.9 | 57.5 |
| CMS-LW | 93.3 | 92.5 | 75.4 | 59.2 |
| ACW+σ-LPCC | 88.6 | 96.3 | 36.3 | 29.6 |
| PF | 93.3 | 97.1 | 45 | 37.1 |
| CMS+PF | 93.3 | 92.5 | 75.8 | 60 |
| CMS+PF+σ-LPCC | 92.9 | 93.8 | 72.9 | 57.9 |

*Table 8 Identification rates for different training and testing languages, CM*

| PARAMETERIZ. | S4sM1 S4cM1 | S4cM1 S4sM1 | S2sM1 S4cM3 | S2cM2 S4sM3 |
|---|---|---|---|---|
| LPCC | 0.40 2.76 | 1.76 7.53 | 10.40 27.35 | 13.17 19.14 |
| LPCC$_{3..P}$ | 1.50 2.32 | 2.90 13.48 | 11.27 20.59 | 10.04 20.00 |
| σ-LPCC | 1.28 2.92 | 2.37 6.76 | 18.06 22.33 | 18.28 24.87 |
| ACW | 0.27 2.09 | 1.21 6.94 | 10.04 17.02 | 12.04 19.36 |
| CMS | 0.98 2.19 | 1.67 7.97 | 5.25 13.36 | 5.87 14.02 |
| CMS+ACW | 0.81 2.20 | 1.58 8.14 | 4.04 12.91 | 5.14 13.60 |
| CMS+ACW+σ-LPCC | 4.99 4.30 | 1.47 7.53 | 4.24 12.81 | 5.64 12.96 |
| CMS+σ-LPCC | 0.98 2.26 | 2.31 7.82 | 5.88 12.94 | 6.99 12.79 |
| CMS-LW | 0.98 2.19 | 1.67 7.97 | 5.25 13.36 | 5.87 14.02 |
| ACW+σ-LPCC | 1.10 2.20 | 1.96 5.56 | 17.13 21.74 | 17.15 24.26 |
| PF | 0.40 2.76 | 1.76 7.53 | 10.70 17.59 | 12.53 19.28 |
| CMS+PF | 0.98 2.19 | 1.67 7.97 | 5.10 13.90 | 5.68 14.38 |
| CMS+PF+σ-LPCC | 0.98 2.26 | 2.31 7.82 | 5.83 13.32 | 6.43 13.07 |

*Table 9: EER (%) (with cohorts = 5 / without) for different training and testing languages.*

## 4. CONCLUSIONS

In this paper we have studied the relevance of several mismatch conditions in a speaker identification application. It is important because in real applications it is quite difficult to obtain the same conditions in train and test phases (different kinds of microphones, test sentences are recorded in different days than the training sentences, for bilingual speakers it is quite common to change from one language to the other, etc.). Our study includes an exhaustive study of several parameterizations and combinations between them in order to obtain the most robust features to mismatch conditions in all the scenarios. The results of section 5 let us to establish the following conclusions:

- The change of microphone between training and testing sessions is more important than the mismatch of recording session and the mismatch of recording languages.
- The parameterization algorithm (implemented preprocessing algorithm over the LPCC coefficients) is more relevant than the change of the identification algorithm. In fact, the reported differences in identification rates against the change of identification algorithm that can be found in the literature are smaller than the change of the parameterization.
- Although it is possible to find an optimal parameterization for each particular condition, the best parameterization in all the scenarios seems to be the CMS+ACW+σ-LPCC for the VQ algorithm, and CMS+ACW for the CM. For the VQ identification algorithm it obtains a decrease of 1.7% in the identification rates over the classical LPCC parameterization (no mismatch) and an increase of 57.1% for the change of microphone, 20.8% for simultaneous change of microphone and recording session, and 53.7% for the change of language, recording session, and microphone.
- PF and ACW are good parameterizations for "soft" mismatching (see first columns of tables 1 to 9).